\documentclass[aps,prb,twocolumn,showpacs,superscriptaddress]{revtex4}
\usepackage{graphicx,amssymb,amsmath}
\usepackage{natbib}
\usepackage{bm}
\newcommand\CuFeGeO{Cu$_2$Fe$_2$Ge$_4$O$_{13}$}
\newcommand\CuScGeO{Cu$_2$Sc$_2$Ge$_4$O$_{13}$}

\begin{document}
\title{Spin dimers under staggered and random field in \CuFeGeO\ }
\author{T. Masuda}
\affiliation{Department of Nanosystem Science,
Yokohama City University, Yokohama, Kanagawa, 236-0027, Japan}
\email[]{tmasuda@yokohama-cu.ac.jp}

\author{K. Kakurai}
\affiliation{Quantum Beam Science Division, JAEA, Tokai, Ibaraki 319-1195,
Japan}

\author{A. Zheludev}
 \affiliation{Institute for Solid State Physics, ETH, CH-8093
Z\"{u}rich, Switzerland}
 \affiliation{Laboratory for Neutron
Scattering, ETH and Paul Scherrer Institut, CH-5232 Villigen PSI,
Switzerland}
 \altaffiliation{Condensed Matter Science Division, Oak
Ridge National Laboratory, Oak Ridge, TN 37831-6393, USA}

\date{\today}

\begin{abstract}
We study $S=1/2$ dimer excitation in a coupled chain and dimer
compound \CuFeGeO\ by inelastic neutron scattering technique.
The Zeeman split of the dimer triplet by
a staggered field is observed at low temperature.
With the increase of temperature the effect of random field
is detected by a drastic broadening of the triplet excitation.
Basic dynamics of dimer in the staggered and random fields
are experimentally identified in \CuFeGeO .
\end{abstract}


\maketitle


Excitations in quantum spin liquids can be viewed as strongly
interacting bosonic quasiparticles. This circumstance enables
experimental studies of the physics of Bose liquids in prototypical
quantum magnetic
materials\cite{GiamarchiPRB99,Nikuni00,GiamarchiNaturePhys}. Such
experiments are often possible under conditions that can not be
realized in more conventional models, such as
$^4$He\cite{London,Reppy} and ultracold trapped
ions\cite{MHanderson,Wynar}. One recent topic of interest is teh
behavior of bosonic quasiparticles in the presence of disorder.
Exotic new phases such as the Random Singlet state\cite{Ma79}, Bose
and Mott glasses\cite{Fisher89} have been predicted for systems with
quenched disorder. In real prototype materials one usually tries to
create such disorder by chemical doping\cite{ManakaPRL,OosawaPRB}.
In the present work we demonstrate an alternative approach: a random
magnetic field created by disordered (paramagnetic) ions. We show
that such a random field acting on a simple dimer-based quantum spin
liquid dramatically alters the excitation spectrum.

\begin{figure}
\begin{center}
\includegraphics[width=8.5cm]{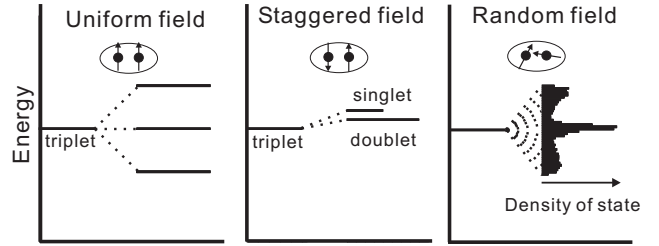}
\end{center}
\caption{ Schematic diagrams of triplet excitations in $S=1/2$
dimers in different types of locally applied magnetic field. }
\label{fig5}
\end{figure}

Let us consider the effect of different types of magnetic fields on
an isolated $S=1/2$ dimer, as shown in Fig.~\ref{fig5}. In a uniform
field the excited triplet is split into three levels. Eventually, at
high field, $\lvert S=1, S^z=0\rangle $ will cross the singlet
ground state. In the presence of inter-dimer interactions, BEC of
magnon will occur. If local fields applied to each dimer spin are
antiparallel to each other (referred to as ``staggered field''
hereafter) the triplet is split into a singlet and a doublet.  The
singlet ground state becomes mixed with $\lvert S=1, S^z=0\rangle $
and the total spin is no longer a good quantum number. Even in an
infinitesimal staggered field the ground state becomes ploarized.
Now, if the field direction is spatially randomized, each dimer will
experience both a staggered and uniform component. The corresponding
energy levels can be calculated numerically. The resulting density
of state (DOS) for excitations in a set of $N$ dimers is plotted in
the right panel in Fig.~\ref{fig5}. The DOS  lower and higher
boundaries of the DOS spread coincide with the levels of $\lvert
S_{z}=1\rangle $ and $\lvert S_{z}=-1\rangle $ in the uniform field.

The quantum ferrimagnet \CuFeGeO\ \cite{Masuda03} is a rare
potential realization  of this random field effect. The compound
includes $S$ = 1/2 Cu$^{2+}$ dimers coupled to classical Fe$^{3+}$
chains \cite{Masuda05}. At low temperature the cooperative ordered
state with classical spin and quantum spin is stabilized by a weak
inter-subsystem coupling. In the adiabatic approximation, the
quantum spins are effectively under the internal field from the much
slower fluctuating classical spins. In this compound, the staggered
nature of the exchange field is due to the magnetic structure. The
staggered magnetization curves of dimers in \CuFeGeO
~\cite{Masuda04a} were experimentally obtained by measuring the
temperature dependence of sublattice moments in neutron diffraction.
At high temperature, in the paramagnetic phase, the classical spins
are thermally disordered and the effective field on the quantum
spins is randomly oriented. Then the system can be considered as the
ensemble of $N$ dimers in a random quasi-static field. As shown in
Fig.~\ref{fig5} the effect of this random field is to broaden the
dimer excitations at $T > T_{\rm N}$.

In the previous inelastic neutron scattering study it was shown that
the energy scales of excitations in the Fe chains and Cu dimers are
well separated~\cite{Masuda05,Masuda07a}. The lower energy
excitations up to 10 meV are Fe-based spin waves. Preliminary powder
experiments~\cite{Masuda05} and comparative studies in an
isostructural compound \CuScGeO ~\cite{Masuda06} associated the
dispersionless excitations at 24 meV  with Cu-dimers. However the
effect of a staggered and/or fluctuating field could not been
identified in powders samples. In the present paper we study the
dimer excitations by single crystal inelastic neutron scattering. By
adopting a high resolution setup, we identify the split peaks due to
the staggered exchange field. Furthermore, we observe a drastic
broadening of the peak profile at $T > T_{\rm N}$ that can be
ascribed to randomly oriented field from thermally fluctuated Fe
moments.

High quality single crystals were grown by floating zone method. The
crystal (monoclinic $P2_1/m$) were found to be twinned, so that both
microscopic domains share $a^*$ - $b^*$ plane. To avoid
complications due to twinning, we restrict the measurements to the
$a^*$ - $b^*$ plane. In the setups Ia and Ib PG (002) were used for
both monochromator and analyzer. The Soller collimations were 48' -
60' - 60' - 120' and open - 80' - 80' - open for Ia and Ib,
respectively. In setup II, to achieve high energy resolution, PG
(004) for monochromator and PG (002) for analyzer with 30' - 20' -
40' - 120' were used. The setups Ia and II were performed on HB1
spectrometer in HFIR, ORNL. The setup Ib was performed on TAS1
spectrometer in JRR-3M, JAEA. In all setups final energy of the
neutron was fixed at $E_f$ = 14.7meV and PG filter was installed
after the sample to eliminate higher order contamination. A closed
cycle He refrigerator was used to achieve low temperatures.

\begin{figure}
\begin{center}
\includegraphics[width=8.7cm]{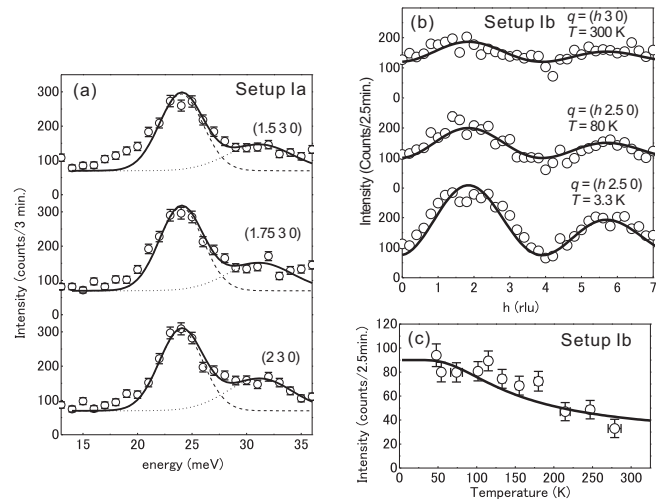}
\end{center}
\caption{ Inelastic neutron scattering using experimental setup Ia
and Ib. (a) Typical energy scans at $(h~k~0)$. Dispersionless
excitations are observed at $\hbar \omega = 24$ and 31 meV. Two
peaks are separately fit by Gaussians (dotted curves). (b) $h$ scans
at $\hbar \omega =$ 24 meV at various temperatures. Sinusoidal
intensity modulations are fitted to the dimer structure factor
calculated for zero field, plus a constant background (solid
curves). (c) Temperature dependence of the peak intensity at ${\bm
q} = (h~2.5~0)$ and $\hbar \omega =$ 24 meV. } \label{fig1}
\end{figure}

In a series of energy scans in a wide range of $(h~k~0)$ space shown
in Fig.\ref{fig1}(a) two dispersionless peaks are readily
identified: a pronounced one at $\hbar \omega \sim 24$ meV and a
weaker feature at $\hbar \omega \sim 31$ meV. The experiments were
performed in setups Ia and Ib. The former is consistent with the
Cu-centered magnetic excitation in previous
studies~\cite{Masuda05,Masuda06}. Constant energy scan at $\hbar
\omega = 24$ meV and its temperature dependence are shown in
Fig.\ref{fig1}(b). The observed sinusoidal intensity modulation is
characteristics of dimer excitations and is observed in a wide
temperature range. In fig.~\ref{fig1}(c) the temperature dependence
of the peak intensity is shown. The intensity at ${\bm q} =
(0~2.5~0)$ was measured at each temperature and then subtracted as
background. The decrease of the intensity at high temperature is
common behavior for magnetic excitations in local spin clusters. The
smaller peak at $\hbar \omega \sim 31$ meV was identified as a
Fe-centered excitation, as will be discussed below.

\begin{figure}
\begin{center}
\includegraphics[width=7.5cm]{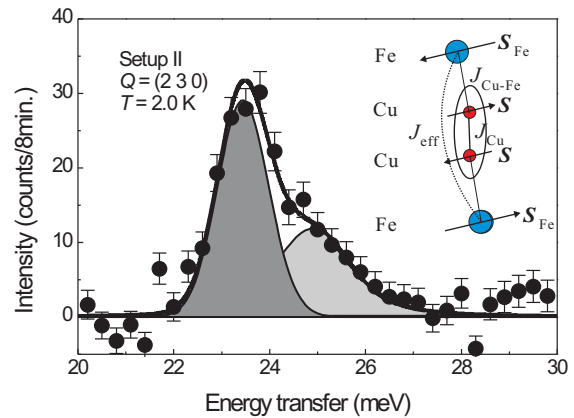}
\end{center}
\caption{ Energy scan collected using the high resolution setup II
at $T$ = 2.0 K. The shoulder structure is reproduced by the doublet
(shaded with white background) and singlet (shaded with gray
background) dimer excitations, split by a staggered exchange field.
} \label{fig3}
\end{figure}

To obtain a more detailed profile, we performed an energy scan using
setup II at $T$ = 2.0 K. As shown in Fig.~\ref{fig3} it is revealed
that the primary peak at $\hbar \omega = 24$ meV actually has a
shoulder structure. The main peak is located at 23.5 meV, and a
smaller bump is centered around 25.0 meV. This splitting is
attributed to the staggered exchange field from the adjacent Fe
moments. The main peak corresponds to the excitation doublet and the
small one to the singlet.

\begin{figure}
\begin{center}
\includegraphics[width=8.7cm]{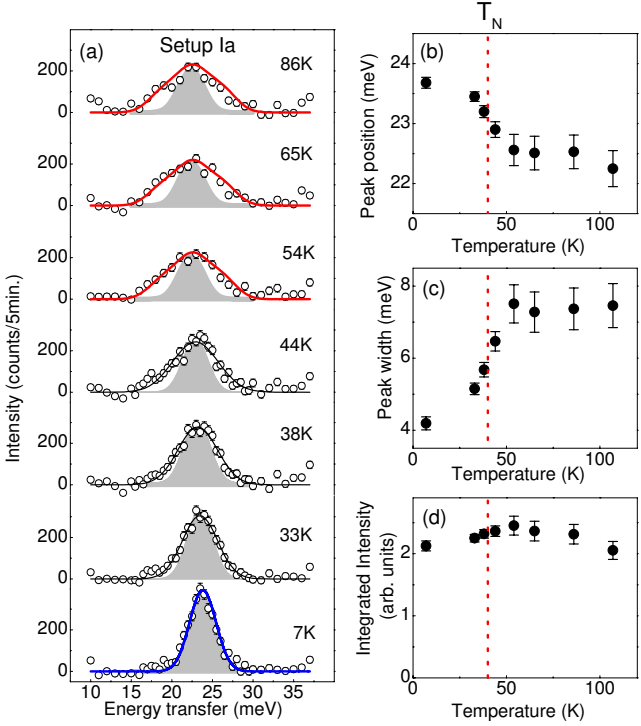}
\end{center}
\caption{ (a) Temperature dependence of the peak profile at ${\bm q}
= (2~3~0)$. Experimental resolution is indicated by the gray area.
Small peaks due to Fe-centered excitation at 31 meV are separately
fitted and subtracted. Profiles at $T \ge 54$ K are reproduced by
dimers in a randomly oriented field model (solid curves). The
temperature dependence of peak positions (b), widths (c) and
integrated intensities (d), as estimated from Gaussin fits. }
\label{fig4}
\end{figure}

Energy scans collected at several temperatures are shown
Fig.~\ref{fig4}(a). The small peak at $\hbar \omega \sim 31$ meV in
Fig.~\ref{fig1} is temperature independent and has been  subtracted
from the data. Well-defined peaks are observed at all temperatures.
While at low temperature the peak profile is sharp and the width is
within resolution limit, at $T \gtrsim T_N$ the peak becomes
drastically broadened. This qualitative behavior is consistent with
the effect a random exchange field should have on the dimer
excitation triplet. The data were analyzed using Gaussian fits. The
estimated peak positions, widths, and the integrated intensities are
plotted as functions of temperature in Figs.~\ref{fig4}(b)-(d). With
increasing temperature the peak energy decreases at $T \sim T_{\rm
N}$,  and stays constant beyond. The peak width drastically
increases at $T \sim T_{\rm N}$, but also remains constant at higher
temperature. The integrated intensity decreases by 10 $\sim $ 20\%.
It is noted that in the previous powder experiment the peak cannot
be distinguished at $T \ge 41$ K \cite{Masuda05}. This is because
the powder integration in wide ${\bm q}$ space collects phonon
excitations and accidental suprious peaks, masking magnetic
excitations at higher temperatures.

For $T < T_{\rm N}$ we will consider the following effective
Hamiltonian:
\begin{equation}
H = J_{\rm Cu}{\bm S_1}\cdot {\bm S_2} +g\mu _{\rm B}{\bm S_1}\cdot {\bm h_1} + g\mu _{\rm B}{\bm S_2}\cdot {\bm h_2},
\label{dimerHamiltonian}
\end{equation}
where ${\bm h_1}=(0~0~h)$ and ${\bm h_2}=(0~0~-h)$. Here  the
$z$-axis is chosen along the ordered Cu moment. The ground state
energy $E_{\rm G}$ decreases with the field, $E_{\rm G} =-J_{\rm
Cu}/4-\sqrt {(2g\mu _{\rm B} h)^2+J_{\rm Cu}^2}/2$, and the
excitation triplet splits into a singlet and doublet. The
corresponding energy levels are giben by
\begin{eqnarray}
\Delta_{\rm s}&=&\sqrt {(2g\mu _{\rm B} h)^2+J_{\rm Cu}^2}
\label{singlet} \\
\Delta_{\rm d}&=&J_{\rm Cu}/2+\sqrt{(2g\mu _{\rm B} h)^2+
J_{\rm Cu}^2}/2, \label{doublet}
\end{eqnarray}
and plotted vs. $h$ in Fig.~\ref{fig5}. For $g\mu _{\rm B}  h\ll
\Delta$,  the neutron cross section is approximately given by:
\begin{eqnarray}
\frac{d^2\sigma}{d\Omega dE}\sim N(\gamma r_0)^2\frac{k'}{k}
\sin ^2(\bm {q.d})\left( f(q)\right) ^2P(T) \nonumber \\
\times \{A(h)(1+\cos ^2 \theta)\delta (\hbar \omega-\Delta_{\rm d})+
B(h)\sin ^2 \theta
\delta(\hbar \omega -\Delta _{\rm s})\}. \label{crosssection}
\end{eqnarray}
The doublet and singlet terms correspond to transverse and
longitudinal spin fluctuation, respectively. $A(h)$ and $B(h)$ are
$h$ dependent parameters with $A(h) \le 1, B(h) \le 1,$ and $A(0) =
B(0) = 1$. Since the staggered field stabilizes the polarized spin
configuration and suppresses longitudinal fluctuation, $B(h)$
decrease with $h$. Meanwhile $A(h)$ is almost constant in the low
field. $P(T)$ is a temperature factor, $P(T)=1/\{ 1+2\exp (-\beta
\Delta _{\rm d}) + \exp (-\beta \Delta _{\rm s}\}$. $\bm q$ is the
scattering vector, $\bm d$ is the spin separation in each dimer, and
$\theta$ is the angle between $\bm q$ and the moment of Cu. We used
${\bm m}_{\rm Cu}$ = (-0.227, 0.035, -0.301) $\mu _{\rm
B}$\cite{Masuda04a} to calculate $\theta$. Two types of domains,
namely antiferromagntic and crystallographic ones due to twinning,
are considered.

The peak profile in Fig.~\ref{fig3} is reasonably well reproduced by
the cross section convoluted by experimental resolution function
with $\Delta _{\rm s}$ = 25.0 meV and $\Delta _{\rm d}$ = 23.5 meV.
From eqs.(\ref{singlet}) and (\ref{doublet}), $J_{\rm Cu}=22.0$ meV
and $h$ = 51 T are obtained. Let us check the consistency of $h$
with the previous study~\cite{Masuda04a}. From the staggered
magnetization curve by neutron diffraction $J_{\rm Cu-Fe}/J_{\rm
Cu}=0.105$ was obtained. Here $J_{\rm Cu-Fe}$ is the interaction
between Cu and Fe spins. Using the molecular field relation
$h=m_{\rm Fe}J_{\rm Cu-Fe}/(g\mu _{\rm B})^2$ and previously
obtained parameters, $h \sim 40$ T is estimated. Thus statically
estimated value is consistent with that obtained in the present
dynamic measurement.

The energy splitting between the singlet and doublet states is about
1.5 meV. This value is small compared energy resolution in the
typical experimental setup. In setup Ia and Ib at $T < T_{\rm N}$,
therefore, staggered field effect is smeared and two terms in
eq.(\ref{crosssection}) are integrated. Then the cross section is
approximately equivalent to that at $h$ = 0. Indeed, the constant
energy scan at $T$ = 3.3 K in Fig.~\ref{fig1}(b) is reasonably
fitted by dimers cross section in zero field shown by the thick
curve.

At $T>T_{\rm N}$ effective field on the Cu dimers is randomly
oriented. We consider an ensemble of $N$ dimers in random field. The
randomly oriented field ${\bm h}$ is assumed to have a constant
magnitude in eq.(\ref{dimerHamiltonian}). The resulting DOS of the
excited states is then calculated numerically. The neutron cross
section is assumed to be approximately proportional to the DOS,
\begin{equation}
\frac{d^2\sigma}{d\Omega dE}=(\gamma r_0)^2\frac{k'}{k}\sin ^2(\bm {q.d})f(q)^2P_{\rm rand}(T) D(\hbar \omega)  \label{crosssection2}
\end{equation}
with $\int D(\epsilon )d\epsilon = 3N$ and $P_{\rm rand}(T) =
N/(N+\int D(\epsilon )e^{-\frac{\epsilon }{k_{\rm B}T}}d\epsilon )$.
The data collected at $T \ge 54$ K are well reproduced by this cross
section convoluted by experimental resolution function, as indicated
by solid curves in Fig.~\ref{fig4}(a). The obtained fit parameters
are $J_{\rm Cu}$ = 22.3(4) meV and $h$ = 41.(8) T. The values are
reasonably consistent with those obtained at $T \le T_{\rm N}$. The
${\bm q}$ dependence of the cross section is the same as for zero
field, and is given by dimer structure factor $\sin ^2 (\bm {q.d})$.
Indeed, the ${\bm q}$ scans at 80 K and 300 K in Fig.~\ref{fig1}(b)
are reproduced by this model. The temperature dependence in
Fig.~\ref{fig1}(c) is well accounted for by the temperature factor
$P_{\rm rand}(T)$.

We shall now discuss the small decrease of the intensity at $T <
T_{\rm N}$ in Fig.~\ref{fig4}(d). At $T > T_{\rm N}$ dimer spins are
fluctuated equally in all directions and the dynamical spin
correlation is fully detected by neutron. In the ordered state the
polarized magnetic ordering suppresses the longitudinal fluctuation
of Cu spins. To estimate the reduction of the longitudinal
excitation we calculate $B(h = 51 {\rm T})=0.77$. The reduction of
$B(h)$ is about 20\% that is consistent with the experiment. This
means that 51 T is rather modest compared with the intradimer
interaction $J_{\rm Cu}$ = 22 meV. If the effective field was large
and the moment were fully polarized, the suppression would be more
drastic. Such a situation is in fact realized in Haldane spin chains
coupled to rare earth moment in Pr$_2$BaNiO$_5$ with fully saturated
Ni$^{2+}$ moment at $T < T_{\rm N}$~\cite{Zheludev96c}. The
Haldane-gap mode lost half of its intensity at $T < T_{\rm N}$ and
it was ascribed to the total suppression of longitudinal mode.

Finally we will mention the temperature independent small peak at
$\hbar \omega \sim$ 31 meV in Fig.~\ref{fig1}(a). If the Fe $S=5/2$
chains were perfectly isolated from the Cu subsystem, the Fe
excitation spectrum  would be dominated by one-magnon excitation at
$\hbar \omega \le 5J_{\rm Fe}$. However, a recent theory predicts
that the introduction of Cu dimer enhances the multi-magnon
excitation of Fe spins at $\hbar \omega$ = $10J_{\rm Fe}$, $15J_{\rm
Fe}$, $20J_{\rm Fe}$ and $25J_{\rm Fe}$. According to the Bond
operator method~\cite{Matsumoto04,Sachdev90} the excitation at
$\hbar \omega$ = $20J_{\rm Fe}$ is the particularly
enhanced~\cite{Matsumoto09}. Since $J_{\rm Fe}=1.6$
meV~\cite{Masuda07a}, the observed small peak at $\hbar \omega$ = 31
meV could be ascribed to the Fe centered longitudinal excitation.
Further details will be published somewhere else.

To conclude, we have experimentally investigated the dynamics of $S
= 1/2$ dimers in staggered and random fields in \CuFeGeO . The
staggered field is realized at $T < T_{\rm N}$ and produces a
splitting of the excitation triplet.  At $T > T_{\rm N}$ a random
exchange field produces a drastic broadening of these modes. In teh
future, polarized neutron experiments may be useful to separate the
longitudinal and transverse excitations. Recently
Cu$_2$CdB$_2$O$_6$~\cite{Hase05} and
Cu$_3$Mo$_2$O$_9$~\cite{Hamasaki08} identified as new realizationsof
the coupled dimers and chains models. Particularly in the latter
compound, the dimer energy is close to that of the  chains, and more
complex physics is expected.

Prof. M. Matsumoto is greatly appreciated for fruitful discussion.
This work was partly supported by Yamada Science Foundation, Asahi
glass foundation, and Grant-in-Aid for Scientific Research (No.s
19740215 and 19052004) of Ministry of Education, Culture, Sports,
Science and Technology of Japan.


\begin{thebibliography}{22}
\expandafter\ifx\csname natexlab\endcsname\relax\def\natexlab#1{#1}\fi
\expandafter\ifx\csname bibnamefont\endcsname\relax
  \def\bibnamefont#1{#1}\fi
\expandafter\ifx\csname bibfnamefont\endcsname\relax
  \def\bibfnamefont#1{#1}\fi
\expandafter\ifx\csname citenamefont\endcsname\relax
  \def\citenamefont#1{#1}\fi
\expandafter\ifx\csname url\endcsname\relax
  \def\url#1{\texttt{#1}}\fi
\expandafter\ifx\csname urlprefix\endcsname\relax\def\urlprefix{URL }\fi
\providecommand{\bibinfo}[2]{#2}
\providecommand{\eprint}[2][]{\url{#2}}

\bibitem[{\citenamefont{Giamarchi and Tsvelik}(1999)}]{GiamarchiPRB99}
\bibinfo{author}{\bibfnamefont{T.}~\bibnamefont{Giamarchi}} \bibnamefont{and}
  \bibinfo{author}{\bibfnamefont{A.~M.} \bibnamefont{Tsvelik}},
  \bibinfo{journal}{Phys. Rev. B} \textbf{\bibinfo{volume}{59}},
  \bibinfo{pages}{11398} (\bibinfo{year}{1999}).

\bibitem[{\citenamefont{Nikuni et~al.}(2000)\citenamefont{Nikuni, Oshikawa,
  Oosawa, and Tanaka}}]{Nikuni00}
\bibinfo{author}{\bibfnamefont{T.}~\bibnamefont{Nikuni}},
  \bibinfo{author}{\bibfnamefont{M.}~\bibnamefont{Oshikawa}},
  \bibinfo{author}{\bibfnamefont{A.}~\bibnamefont{Oosawa}}, \bibnamefont{and}
  \bibinfo{author}{\bibfnamefont{H.}~\bibnamefont{Tanaka}},
  \bibinfo{journal}{Phys. Rev. Lett.} \textbf{\bibinfo{volume}{84}},
  \bibinfo{pages}{5868} (\bibinfo{year}{2000}).

\bibitem[{\citenamefont{Giamarchi et~al.}(2008)\citenamefont{Giamarchi, Ruegg,
  and Tchernyshev}}]{GiamarchiNaturePhys}
\bibinfo{author}{\bibfnamefont{T.}~\bibnamefont{Giamarchi}},
  \bibinfo{author}{\bibfnamefont{C.}~\bibnamefont{Ruegg}}, \bibnamefont{and}
  \bibinfo{author}{\bibfnamefont{O.}~\bibnamefont{Tchernyshev}},
  \bibinfo{journal}{Nature Physics} \textbf{\bibinfo{volume}{4}},
  \bibinfo{pages}{198} (\bibinfo{year}{2008}).

\bibitem[{\citenamefont{London}(1938)}]{London}
\bibinfo{author}{\bibfnamefont{F.}~\bibnamefont{London}},
  \bibinfo{journal}{Nature} \textbf{\bibinfo{volume}{141}},
  \bibinfo{pages}{643} (\bibinfo{year}{1938}).

\bibitem[{\citenamefont{Reppy and Depatte}(1964)}]{Reppy}
\bibinfo{author}{\bibfnamefont{J.~D.} \bibnamefont{Reppy}} \bibnamefont{and}
  \bibinfo{author}{\bibfnamefont{D.}~\bibnamefont{Depatte}},
  \bibinfo{journal}{Phys. Rev. Lett.} \textbf{\bibinfo{volume}{12}},
  \bibinfo{pages}{187} (\bibinfo{year}{1964}).

\bibitem[{\citenamefont{Anderson et~al.}(1995)\citenamefont{Anderson, Ensher,
  Matthews, Wieman, and Cornell}}]{MHanderson}
\bibinfo{author}{\bibfnamefont{M.~H.} \bibnamefont{Anderson}},
  \bibinfo{author}{\bibfnamefont{J.~R.} \bibnamefont{Ensher}},
  \bibinfo{author}{\bibfnamefont{M.~R.} \bibnamefont{Matthews}},
  \bibinfo{author}{\bibfnamefont{C.~E.} \bibnamefont{Wieman}},
  \bibnamefont{and} \bibinfo{author}{\bibfnamefont{E.~A.}
  \bibnamefont{Cornell}}, \bibinfo{journal}{Science}
  \textbf{\bibinfo{volume}{269}}, \bibinfo{pages}{198} (\bibinfo{year}{1995}).

\bibitem[{\citenamefont{Wynar et~al.}(2000)\citenamefont{Wynar, Freeland, Han,
  Ryu, and Heinzen}}]{Wynar}
\bibinfo{author}{\bibfnamefont{R.}~\bibnamefont{Wynar}},
  \bibinfo{author}{\bibfnamefont{R.~S.} \bibnamefont{Freeland}},
  \bibinfo{author}{\bibfnamefont{D.~J.} \bibnamefont{Han}},
  \bibinfo{author}{\bibfnamefont{C.}~\bibnamefont{Ryu}}, \bibnamefont{and}
  \bibinfo{author}{\bibfnamefont{D.~J.} \bibnamefont{Heinzen}},
  \bibinfo{journal}{Science} \textbf{\bibinfo{volume}{287}},
  \bibinfo{pages}{1016} (\bibinfo{year}{2000}).

\bibitem[{\citenamefont{k.~Ma et~al.}(1979)\citenamefont{k.~Ma, Dasgupta, and
  k.~Hu}}]{Ma79}
\bibinfo{author}{\bibfnamefont{S.}~\bibnamefont{-k.~Ma}},
  \bibinfo{author}{\bibfnamefont{C.}~\bibnamefont{Dasgupta}}, \bibnamefont{and}
  \bibinfo{author}{\bibfnamefont{C.}~\bibnamefont{-k.~Hu}},
  \bibinfo{journal}{Phys. Rev. Lett.} \textbf{\bibinfo{volume}{43}},
  \bibinfo{pages}{1434} (\bibinfo{year}{1979}).

\bibitem[{\citenamefont{Fisher et~al.}(1989)\citenamefont{Fisher, Weichman,
  Grinstein, and Fisher}}]{Fisher89}
\bibinfo{author}{\bibfnamefont{M.~P.~A.}~\bibnamefont{Fisher}},
  \bibinfo{author}{\bibfnamefont{P.~B.}~\bibnamefont{Weichman}},
  \bibinfo{author}{\bibfnamefont{G.}~\bibnamefont{Grinstein}},
  \bibnamefont{and} \bibinfo{author}{\bibfnamefont{D.~S.}~\bibnamefont{Fisher}},
  \bibinfo{journal}{Phys. Rev. B} \textbf{\bibinfo{volume}{40}},
  \bibinfo{pages}{546} (\bibinfo{year}{1989}).

\bibitem[{\citenamefont{Manaka et~al.}(2008)\citenamefont{Manaka, Kolomietz,
  and Goto}}]{ManakaPRL}
\bibinfo{author}{\bibfnamefont{H.}~\bibnamefont{Manaka}},
  \bibinfo{author}{\bibfnamefont{A.~V.} \bibnamefont{Kolomiets}},
  \bibnamefont{and} \bibinfo{author}{\bibfnamefont{T.}~\bibnamefont{Goto}},
  \bibinfo{journal}{Phys. Rev. Lett.} \textbf{\bibinfo{volume}{101}},
  \bibinfo{pages}{077204} (\bibinfo{year}{2008}).

\bibitem[{\citenamefont{Oosawa and Tanaka}(2002)}]{OosawaPRB}
\bibinfo{author}{\bibfnamefont{A.}~\bibnamefont{Oosawa}} \bibnamefont{and}
  \bibinfo{author}{\bibfnamefont{H.}~\bibnamefont{Tanaka}},
  \bibinfo{journal}{Phys. Rev. B} \textbf{\bibinfo{volume}{65}},
  \bibinfo{pages}{184437} (\bibinfo{year}{2002}).

\bibitem[{\citenamefont{Masuda et~al.}(2003)\citenamefont{Masuda, Chakoumakos,
  Nygren, Imai, and Uchinokura}}]{Masuda03}
\bibinfo{author}{\bibfnamefont{T.}~\bibnamefont{Masuda}},
  \bibinfo{author}{\bibfnamefont{B.~C.} \bibnamefont{Chakoumakos}},
  \bibinfo{author}{\bibfnamefont{C.~L.} \bibnamefont{Nygren}},
  \bibinfo{author}{\bibfnamefont{S.}~\bibnamefont{Imai}}, \bibnamefont{and}
  \bibinfo{author}{\bibfnamefont{K.}~\bibnamefont{Uchinokura}},
  \bibinfo{journal}{J. Solid State Chem.} \textbf{\bibinfo{volume}{176}},
  \bibinfo{pages}{175} (\bibinfo{year}{2003}).

\bibitem[{\citenamefont{Masuda et~al.}(2005)\citenamefont{Masuda, Zheludev,
  Sales, Imai, Uchinokura, and Park}}]{Masuda05}
\bibinfo{author}{\bibfnamefont{T.}~\bibnamefont{Masuda}},
  \bibinfo{author}{\bibfnamefont{A.}~\bibnamefont{Zheludev}},
  \bibinfo{author}{\bibfnamefont{B.}~\bibnamefont{Sales}},
  \bibinfo{author}{\bibfnamefont{S.}~\bibnamefont{Imai}},
  \bibinfo{author}{\bibfnamefont{K.}~\bibnamefont{Uchinokura}},
  \bibnamefont{and} \bibinfo{author}{\bibfnamefont{S.}~\bibnamefont{Park}},
  \bibinfo{journal}{Phys. Rev. B} \textbf{\bibinfo{volume}{72}},
  \bibinfo{pages}{094434} (\bibinfo{year}{2005}).

\bibitem[{\citenamefont{Masuda et~al.}(2004)\citenamefont{Masuda, Zheludev,
  Grenier, Imai, Uchinokura, Ressouche, and Park}}]{Masuda04a}
\bibinfo{author}{\bibfnamefont{T.}~\bibnamefont{Masuda}},
  \bibinfo{author}{\bibfnamefont{A.}~\bibnamefont{Zheludev}},
  \bibinfo{author}{\bibfnamefont{B.}~\bibnamefont{Grenier}},
  \bibinfo{author}{\bibfnamefont{S.}~\bibnamefont{Imai}},
  \bibinfo{author}{\bibfnamefont{K.}~\bibnamefont{Uchinokura}},
  \bibinfo{author}{\bibfnamefont{E.}~\bibnamefont{Ressouche}},
  \bibnamefont{and} \bibinfo{author}{\bibfnamefont{S.}~\bibnamefont{Park}},
  \bibinfo{journal}{Phys. Rev. Lett.} \textbf{\bibinfo{volume}{93}},
  \bibinfo{pages}{077202} (\bibinfo{year}{2004}).

\bibitem[{\citenamefont{Masuda et~al.}(2007)\citenamefont{Masuda, Kakurai,
  Matsuda, Kaneko, and Metoki}}]{Masuda07a}
\bibinfo{author}{\bibfnamefont{T.}~\bibnamefont{Masuda}},
  \bibinfo{author}{\bibfnamefont{K.}~\bibnamefont{Kakurai}},
  \bibinfo{author}{\bibfnamefont{M.}~\bibnamefont{Matsuda}},
  \bibinfo{author}{\bibfnamefont{K.}~\bibnamefont{Kaneko}}, \bibnamefont{and}
  \bibinfo{author}{\bibfnamefont{N.}~\bibnamefont{Metoki}},
  \bibinfo{journal}{Phys. Rev. B} \textbf{\bibinfo{volume}{75}},
  \bibinfo{pages}{220401(R)} (\bibinfo{year}{2007}).

\bibitem[{\citenamefont{Masuda and Redhammer}(2006)}]{Masuda06}
\bibinfo{author}{\bibfnamefont{T.}~\bibnamefont{Masuda}} \bibnamefont{and}
  \bibinfo{author}{\bibfnamefont{G.~J.} \bibnamefont{Redhammer}},
  \bibinfo{journal}{Phys. Rev. B} \textbf{\bibinfo{volume}{74}},
  \bibinfo{pages}{054418} (\bibinfo{year}{2006}).

\bibitem[{\citenamefont{Zheludev et~al.}(1996)\citenamefont{Zheludev,
  Tranquada, Vogt, and Buttrey}}]{Zheludev96c}
\bibinfo{author}{\bibfnamefont{A.}~\bibnamefont{Zheludev}},
  \bibinfo{author}{\bibfnamefont{J.~M.}~\bibnamefont{Tranquada}},
  \bibinfo{author}{\bibfnamefont{T.}~\bibnamefont{Vogt}}, \bibnamefont{and}
  \bibinfo{author}{\bibfnamefont{D.~J.}~\bibnamefont{Buttrey}},
  \bibinfo{journal}{Phys. Rev. B} \textbf{\bibinfo{volume}{54}},
  \bibinfo{pages}{6437} (\bibinfo{year}{1996}).

\bibitem[{\citenamefont{Matsumoto et~al.}(2004)\citenamefont{Matsumoto,
  Normand, Rice, and Sigrist}}]{Matsumoto04}
\bibinfo{author}{\bibfnamefont{M.}~\bibnamefont{Matsumoto}},
  \bibinfo{author}{\bibfnamefont{B.}~\bibnamefont{Normand}},
  \bibinfo{author}{\bibfnamefont{T.~M.}~\bibnamefont{Rice}}, \bibnamefont{and}
  \bibinfo{author}{\bibfnamefont{M.}~\bibnamefont{Sigrist}},
  \bibinfo{journal}{Phys. Rev. B} \textbf{\bibinfo{volume}{69}},
  \bibinfo{pages}{054423} (\bibinfo{year}{2004}).

\bibitem[{\citenamefont{Sachdev and Bhatt}(1990)}]{Sachdev90}
\bibinfo{author}{\bibfnamefont{S.}~\bibnamefont{Sachdev}} \bibnamefont{and}
  \bibinfo{author}{\bibfnamefont{R.~N.}~\bibnamefont{Bhatt}},
  \bibinfo{journal}{Phys. Rev. B} \textbf{\bibinfo{volume}{41}},
  \bibinfo{pages}{9323} (\bibinfo{year}{1990}).

\bibitem{Matsumoto09}
M. Matsumoto, private communication.

\bibitem[{\citenamefont{Hase et~al.}(2005)\citenamefont{Hase, Kohno, Kitazawa,
  Suzuki, Ozawa, and Kido}}]{Hase05}
\bibinfo{author}{\bibfnamefont{M.}~\bibnamefont{Hase}},
  \bibinfo{author}{\bibfnamefont{M.}~\bibnamefont{Kohno}},
  \bibinfo{author}{\bibfnamefont{H.}~\bibnamefont{Kitazawa}},
  \bibinfo{author}{\bibfnamefont{O.}~\bibnamefont{Suzuki}},
  \bibinfo{author}{\bibfnamefont{K.}~\bibnamefont{Ozawa}},
  \bibinfo{author}{\bibfnamefont{G.}~\bibnamefont{Kido}},
  \bibinfo{author}{\bibfnamefont{M.}~\bibnamefont{Imai}},\bibnamefont{and}
  \bibinfo{author}{\bibfnamefont{X.}~\bibnamefont{Hu}},
  \bibinfo{journal}{Phys. Rev. B} \textbf{\bibinfo{volume}{72}},
  \bibinfo{pages}{172412} (\bibinfo{year}{2005}).

\bibitem[{\citenamefont{Hamasaki et~al.}(2008)\citenamefont{Hamasaki, Ide,
  Kuroe, Sekine, Hase, Tsukada, and Sakakibara}}]{Hamasaki08}
\bibinfo{author}{\bibfnamefont{T.}~\bibnamefont{Hamasaki}},
  \bibinfo{author}{\bibfnamefont{T.}~\bibnamefont{Ide}},
  \bibinfo{author}{\bibfnamefont{H.}~\bibnamefont{Kuroe}},
  \bibinfo{author}{\bibfnamefont{T.}~\bibnamefont{Sekine}},
  \bibinfo{author}{\bibfnamefont{M.}~\bibnamefont{Hase}},
  \bibinfo{author}{\bibfnamefont{I.}~\bibnamefont{Tsukada}}, \bibnamefont{and}
  \bibinfo{author}{\bibfnamefont{T.}~\bibnamefont{Sakakibara}},
  \bibinfo{journal}{Phys. Rev. B} \textbf{\bibinfo{volume}{77}},
  \bibinfo{pages}{134419} (\bibinfo{year}{2008}).

\end{thebibliography}
\end{document}